\documentclass[aps,prb,twocolumn,showpacs,superscriptaddress]{revtex4-1}

\usepackage{graphicx} \newcommand{\kv}{\mathbf{k}}
\newcommand{\Rv}{\mathbf{R}} \newcommand{\qv}{\mathbf{q}}
\newcommand{\dv}{{\boldsymbol\delta}}
\newcommand{\ev}{{\boldsymbol\eta}} \newcommand{\x}{\mathbf{x}}
\newcommand{\y}{\mathbf{y}}

\usepackage{amsmath} \newcommand{\EqLabel}[1]{\label{#1}}

\begin{document}
 
\title{Momentum Average approximation for models with boson-modulated
  hopping: the role of closed loops in the dynamical generation of a
  finite quasiparticle mass}

\author{Mona Berciu}

\affiliation{Department of Physics and Astronomy, University of
  British Columbia, Vancouver, BC V6T 1Z1, Canada }

\author{Holger Fehske}

\affiliation{Institut f\"ur Physik, Ernst-Moritz-Arndt-Universit\"at
  Greifswald, D-17487 Greifswald, Germany }

\date{\today}
\begin{abstract}
We generalize the momentum average (MA) approximation to study the
properties of single polarons in models with boson affected hopping,
where the fermion-boson scattering depends explicitly on both the
fermion's and the boson's momentum.  As a specific example, we
investigate the Edwards fermion-boson model in both one- and
two-dimensions. In one dimension, this allows us to compare our
results with Exact Diagonalization results, to validate the accuracy
of our approximation. The generalization to two-dimensional lattices
allows us to calculate the polaron's quasiparticle weight and
dispersion throughout the Brillouin zone, and to demonstrate the
importance of Trugman loops in generating a finite effective mass even
when the free fermion has an infinite mass.
\end{abstract}

\maketitle
\section{Introduction}
One of the most common problems in condensed matter physics is that of
understanding the behavior of a particle coupled to bosons from its
environment, for example an electron interacting with phonons,
magnons, or orbitons of its host
crystal.~\cite{Fr54,Fi75,KLR89,MH91a,WOH09,Be09} The particle becomes
``dressed'' by a cloud of bosonic excitations that accompany it.  The
resulting composite object, generally known as a polaron, can have
properties significantly different from that of the bare
particle.~\cite{La33,IRF06,Alex07}

The theoretical study of these properties is rather difficult away
from the various asymptotic regimes where perturbation theory
holds. Of course, a large variety of numerical techniques have been
developed to deal with such
problems,\cite{JF07,FT07,num1,num2,num3,num4,num5} and many
interesting results have been uncovered, although the focus so far has
been primarily on rather simple models such as the Holstein
Hamiltonian~\cite{Ho59a,Ho59b} that describes the simplest possible
electron-phonon coupling.  The progress on analytical approximations
that can efficiently yet accurately describe the non-perturbative
regimes has been slower. In fact, it is only recently that the
so-called Momentum Average (MA) approximation has been proposed for
the Holstein model, and shown to accurately capture its polaronic
behavior in all the parameter space except the extreme adiabatic
limit.\cite{MA0} A way to systematically improve this approximation,
as well as generalizations to certain kinds of more complex models
have been proposed since.\cite{MA1,MAq,MAp,MAg,MAso,MAd,MAs} The
availability of such simple yet accurate approximations is important,
as it allows one to quickly explore large regions of the parameter
space to identify the interesting properties of the model.

In this work we present the generalization of MA-type methods to
calculate single polaron Green's functions for Hamiltonians whose
hopping is boson affected. The bosons are assumed to be
dispersionless, i.e. of Einstein type.  For most polaron models,
including the one discussed here,
the spin of the fermion is irrelevant and we 
ignore it. Exceptions occur, for example, in systems with spin-orbit
coupling, where suitable generalizations can be implemented.\cite{MAso}
The fermion moves on a $d$-dimensional lattice, which for
simplicity is assumed to be  hypercubic (generalization to other
types of lattices is straightforward\cite{MAg}). The cases $d=1$ and $d=2$
for the Edwards fermion-boson model~\cite{Ed06} are discussed in
detail and interesting physics related to the role of closed loops,
possible in two dimensions (2D) but absent in 1D, is uncovered. We
note that single polaron properties for this model have been
investigated numerically in 1D,~\cite{AEF07} and we use these results
to assess the accuracy of MA. We then extend our method to 2D, where
no results are currently available, and where we illustrate
interesting effects of the boson modulated hopping. Other such models
can be treated similarly.

The general form of the Hamiltonian of interest is:
\begin{eqnarray}
\EqLabel{1} {\cal H}&=& \sum_{\kv}^{} \epsilon_{\kv}c^{\dagger}_{\kv}
        c_{\kv}^{} + \Omega \sum_{\qv}^{} b^\dagger_{\qv}
        b_{\qv}^{}\nonumber\\&& + \sum_{\kv,\qv} \frac{g(\kv,
        \qv)}{\sqrt{N}} \,c^{\dagger}_{\kv-\qv} c_{\kv}^{} \left(
        b^\dagger_{\qv} + b_{-{\qv}}^{}\right)\,.
\end{eqnarray}
Here, $c_\kv$ and $ b_{\qv}$ are fermion, respectively boson
annihilation operators, and $N$ is the number of sites in the
system. In all our results we assume periodic boundary conditions and
let $N\rightarrow \infty$, however finite size systems and/or other
types of boundary conditions can be treated similarly.  The sums are
over the Brillouin zone, $\epsilon_{\kv}$ is the free-fermion
dispersion while $\Omega$ is the bosons' energy (we set
$\hbar=1$). Note that in~\eqref{1} the fermion-boson scattering
depends {\em explicitly} on both the fermion's and the boson's
momentum. This is to be contrasted with simpler cases, such as the
Holstein model, where $g(\kv, \qv) \rightarrow g$ is a constant, or
models where the bosons modulate only on-site energies but not the
hopping integrals, for which $g(\kv, \qv) \rightarrow g(\qv)$. The
accuracy of MA approximations for these simpler types of Hamiltonians
has been demonstrated in Refs. \onlinecite{MA0,MA1,MAq}.

We are interested in calculating the single polaron Green's function,
defined as
\begin{equation}
\EqLabel{2} G(\kv,\omega) = \langle 0| c_{\kv} \hat{G}(\omega)
c^{\dagger}_{\kv}|0\rangle\,,
\end{equation}
where $|0\rangle$ is the vacuum, $\hat{G}(\omega) =[\omega-{\cal
	H}+i\eta]^{-1}$ is the resolvent associated with the Hamiltonian
	${\cal H}$, and $\eta$ is a positive, infinitesimally small
	number.

From the Green's function we get the spectral weight
\begin{equation}
\EqLabel{3} A(\kv,\omega) = -{1\over \pi} \mbox{Im} G(\kv,\omega)
\end{equation}
which is measurable by (Inverse) Angle-Resolved Photoemission
Spectroscopy.\cite{ARPES} The lowest-energy pole of $A(\kv,\omega)$
allows us to 
identify the polaron dispersion $E(\kv)$. Its residue is the
quasiparticle ({\em qp}) weight,
\begin{equation}
\EqLabel{3b} Z_\kv = | \langle \phi_{\kv}|
c^{\dagger}_{\kv}|0\rangle|^2\,,
\end{equation}
i.e. the overlap between the polaron eigenfunction
$|\phi_{\kv}\rangle$, where ${\cal
H}|\phi_{\kv}\rangle=E(\kv)|\phi_{\kv}\rangle$, and a free-fermion
state. Of course, the spectral weight contains information about the
higher energy states as well, but here we will primarily focus on the
low-energy polaron band.

The article is organized as follows. We first introduce the Edwards
fermion-boson model, which is the specific model with boson-modulated
hopping that we will use as an example in this work. We then outline
the MA approximation for the simpler 1D case, and use comparison with
available numerical results to analyze its accuracy in various
regimes. Then, we generalize MA for the 2D case and use it to
understand the relevance of closed loops for generating a dynamical
mass for the dresses quasiparticle. Finally we summarize our results
and conclude.

\section{The Edwards fermion-boson model}

The Edwards fermion-boson model~\cite{Ed06,AEF07} is defined by the
Hamiltonian
\begin{equation}
\EqLabel{4} {\cal H} = -t_f \sum_{\langle i,j\rangle}c^\dagger_i
c_j^{} +\Omega \sum_{i}^{} b^\dagger_i b_i^{} - t_b \sum_{\langle i,
j\rangle}^{} c^\dagger_ic_j(b_j^\dagger + b_i^{})\,,
\end{equation}
where the first term describes nearest-neighbor (NN) hopping of the
fermion on the lattice of interest, the second term describes the
Einstein boson branch, and the last term is the boson-modulated
hopping. Note that in the limit $t_f\rightarrow0$, it is only the last
term that allows the fermion to move: hopping from one site to a
neighboring one either creates an excitation at the ``departure''
site, or removes one from the ``arrival'' site. This model provides a
way to mimic, for example, the motion of a fermion through an
antiferromagnetically ordered spin background.~\cite{KLR89,Tr88,AEF10}
For a Ne\'el antiferromagnet (AFM) doped with one fermion---which is
the zero-order description of a hole moving in a cuprate CuO$_2$
layer---the hopping of the fermion reshuffles the spins along its
path. If the spin at the ``arrival'' site has the proper orientation,
when it is shuffled to the starting site as the fermion hops it will
have the wrong orientation (a ``magnon'' defect is created at the
initial site). Vice versa, if the visited spin has the wrong
orientation, when shuffled by one site it will be properly aligned
(``magnon'' defect removed from the arrival site). This is precisely
the type of physics described by this boson-modulated hopping,
although it ignores details such as the hard-core boson constraint for
magnons, the fact that in a Ne\'el AFM the energy of neighboring
magnons is not additive, and also it allows the particle to coexist
with bosons at the same site. The free-fermion hopping term $t_f$, on
the other hand, has to be added when describing motion through a
Heisenberg-like AFM, where spin fluctuations continuously create and
annihilate magnons. Indeed, as shown in
Refs.~\onlinecite{AEF07,FAW09}, the free-fermion hopping term can be
mapped into a purely bosonic term of the type $\lambda \sum_{i}^{}
(b_i + b_i^\dagger)$ with $\lambda=t_f\Omega/(2t_b)$, which allows the
number of magnons to fluctuate.

\begin{figure}[t]
\includegraphics[width=\columnwidth]{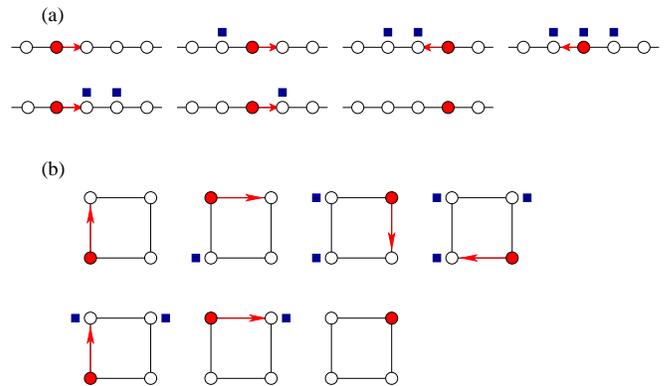}
\caption{Sketch of the 3-boson, 3-site sequence of processes that give
  rise to effective second NN fermion hopping. The site occupied by
  the fermion is shaded and the arrow indicates the direction of the
  next $t_b$-hopping process. Each $t_b$-hopping either leaves a boson
  (drawn as a square) at the initial site, or absorbs a boson from the
  arrival site. Both collinear (a) and closed loop (b) processes are
  allowed in 2D; in 1D only (a) is possible.
\label{fig0}}
\end{figure}

 The ``conventional wisdom'' is that a fermion in a 2D Ne\'el AFM
 ($t_f=0$ case) has an infinite effective mass, because as it tries to
 move it creates a costly string of defects which effectively pin it
 to its original site. This is, however, not true. The boson-modulated
 hopping gives rise to an effective fermion mass even when the bare
 fermionic mass is infinite ($t_f=0$).~\cite{AEF07,AEF10} For the 1D
 chain, this is primarily due to the 3-site, 3-boson processes
 sketched in Fig.~\ref{fig0}(a) which result in an effective second NN
 hopping of the fermion (of course, more complicated processes
 involving more bosons are also possible, but they are energetically
 more costly). In Ref. \onlinecite{Tr88}, it was noted that in 2D, these
 collinear processes---which in 2D give rise to effective 3rd NN
 hopping---are supplemented by the closed loop processes sketched in
 Fig.~\ref{fig0}(b), which give rise to effective 2nd NN hopping. The
 importance of such closed loops---known as Trugman paths---for
 determining the effective quasiparticle mass has been emphasized
 already in Ref.~\onlinecite{Tr88} in the context of cuprates. In this
 work, we are the first to explicitly investigate this phenomenology
 for the 2D Edwards fermion-boson model.

While we focus on this Hamiltonian for the remainder of this work, the
MA method can be generalized straightforwardly to other
boson-modulated hoppings, like that appearing in the 1D
Su-Schrieffer-Heeger model of polyacetylene.~\cite{SSH79} Its
phonon-modulated hopping term is proportional to $\sum_{i}^{}\big(
c^\dagger_i c_{i+1}^{} + {\rm H.c.}\big) \big[b_i^\dagger +
b_i^{}-b_{i+1}^\dagger - b_{i+1}^{} \big]$, i.e. here phonons can be
both absorbed and created at either of the two sites involved in the
hopping process.

\section{The Momentum Average approximation}

One can discuss the meaning of the MA approximations from several
different points of view. One is that this is an approximation which
sums semi-analytically {\em all diagrams} contributing to the
self-energy, however ``exponentially small'' contributions are ignored
when calculating the expression of each such diagram.~\cite{MA1} The precise
meaning of this statement will be clarified below.

A more useful starting point for our purposes here is the variational
meaning of MA.\cite{bat,MA1} The central idea of polaron physics is that the
dressed quasiparticle---the polaron---consists of the fermion itself
and a cloud of bosons in its vicinity. MA permits one to
systematically select what bosonic states to keep in the variational
space to describe this cloud, and sum their contributions
efficiently. Of course, the more states kept, the more accurate the
results. Physical intuition is needed to decide what is a minimal
acceptable starting point.

In this section, we first describe the MA approximation for a 1D
chain, and assess its accuracy against available numerical data. We
then briefly review the 2D generalization and analyze the resulting
physics, which has not been investigated before.

\subsection{MA for the 1D chain}

\begin{figure}[t]
\includegraphics[width=\columnwidth]{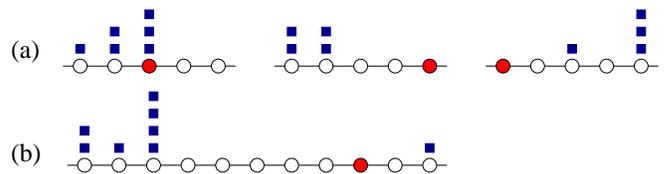}
\caption{(Color online) (a) Sketch of several states included in
  the MA$^{(0)}$ approximation discussed here. The polaron cloud is
  allowed to extend on up to three neighboring sites, located anywhere
  in the system with any number of bosons (shown as  red squares) at
  each site, and the fermion (shown as a shaded blue circle) at any distance
  away from the cloud. (b) Example of extra states included in
  a MA$^{(1)}$ level approximation, with a boson arbitrarily far away
  from the polaron cloud.
\label{fig0b}}
\end{figure}

Since we primarily focus on the low-energy polaronic physics, we will
only attempt to describe the polaronic cloud. The only restriction we
impose regards its spatial size, i.e., what is the maximum number of
neighboring sites that it can span. The number of bosons at any site
in the cloud, as well as the position of the fermion with respect to
the center of the cloud, are not restricted---they can take any
values. A few of the possible states for a 3-site cloud calculation
are sketched in Fig.~\ref{fig0b}(a). Of course, any 1- and 2-site
cloud states belongs to this set.

 Note that here we do not allow any bosonic excitations to occur far
 from the polaron cloud, since they primarily contribute to
 higher-energy states. In other words, this will be a MA$^{(0)}$ level
 approximation.\cite{MA1} Generalization to MA$^{(1)}$ and higher levels,
 needed for example to describe the polaron+one-boson continuum, is
 straightforward although fairly cumbersome. For example, in
 MA$^{(1)}$ we also include states such as sketched in
 Fig.~\ref{fig0b}(b), with one boson arbitrarily far away from the
 polaron cloud. Similarly MA$^{(2)}$ allows two bosons at arbitrary
 locations away from the cloud, etc.  As detailed below, the physics
 encoded in these higher-level approximations does not lead to any
 qualitative changes for the properties and parameter ranges that we
 are interested in.

After deciding on the MA$^{(0)}$ (henceforth called simply MA) level,
the only question left is how big should one allow the polaron cloud
to be. For the Holstein model, a one-site cloud already gives a
remarkably accurate description in any dimension, most everywhere in
the parameter space except intermediate coupling in the adiabatic
limit $\Omega/t\rightarrow 0$.\cite{MA0,MA1}

For the Edwards model, as already discussed, at least 3-site boson
clouds need be considered in order to describe the leading processes
that result in the dynamical generation of a finite effective mass in
the limit $t_f,\,\lambda \rightarrow 0$. As a result, we will start
directly by building a variational MA approximation allowing any
number of bosons on any 3 consecutive sites, which can be located at
any distance from where the fermion is. To achieve this, we introduce
3 types of generalized Green's functions. The first are
\begin{equation}
\EqLabel{4b} F_n(k,q,\omega) = \sum_{i}^{} e^{i(k-q)R_i}\langle 0|
c_k^{} \hat{G}(\omega) c^\dagger_q {b_i^\dagger}^n|0\rangle\,,
\end{equation}
where the ket describes a state of total momentum $k$ (as required
since the Hamiltonian is invariant to translations) of which the
fermion has a momentum $q$ and the cloud of $n$ bosons, all located at
the same site, has momentum $k-q$.

Next are the two-site cloud Green's functions, namely
\begin{equation}
\EqLabel{5} F_{n,m}(k,q,\omega) = \sum_{i}^{} e^{i(k-q)R_i}\langle 0|
c_k \hat{G}(\omega) c^\dagger_q {b_i^\dagger}^m
{b^\dagger}_{i+1}^{n-m}|0\rangle\,,
\end{equation}
which are defined only for $n\ge 2$ and $1\le m\le n-1$, so that they
are distinct from the one-site cloud functions $F_n(k,q,\omega)$
defined above. Finally, we have the three-site cloud functions,
\begin{widetext}
\begin{equation}
\EqLabel{6} F_{n,m,p}(k,q,\omega) = \sum_{i}^{} e^{i(k-q)R_i}\langle
0| c_k \hat{G}(\omega) c^\dagger_q {b^\dagger}_{i-1}^m
{b^\dagger}_{i}^{n-m-p} {b^\dagger}_{i+1}^{p}|0\rangle \,,
\end{equation}
\end{widetext}
defined for $n\ge 2$ and $1\le m\le n-1, 1\le p\le n-1, m+p\le
n$. These restrictions again avoid overlap with the functions
introduced above. Note, however, that they allow bosons to exist only
on the two outer sites of the three-site cloud when $m+p=n$; such
states are not accounted for by $F_{n,m}(k,q,\omega)$. In principle
one can keep adding other states, either with more extended clouds, or
with bosons far away from the cloud, until convergence is
achieved. For reasons already explained, for this model we expect that
it suffices to stop here.

The next step is to generate equations of motion for these generalized
Green's functions. These are obtained by using the Dyson identity
$\hat{G}(\omega) =\hat{G}_0(\omega) +\hat{G}(\omega){\cal V}
\hat{G}_0(\omega)$\,, where $\hat{G}_0(\omega)$ is the resolvent for
${\cal H}_0 = {\cal H}-{\cal V}$. This is an exact (non-perturbative)
identity and holds for any partitioning of ${\cal H}= {\cal H}_0+{\cal
V}$. For our purposes, it is convenient to take ${\cal H}_0$ as the
non-interacting part and ${\cal V}$ as the boson-modulated hopping
term. Note that the kets at the right of $\hat{G}(\omega) $ in all the
above definitions are eigenstates of ${\cal H}_0$, so that, for e.g.
\begin{equation}
\EqLabel{n} \hat{G}_0(\omega) c^\dagger_q {b_i^\dagger}^m
{b^\dagger}_{i+1}^{n-m} |0\rangle = G_0(q, \omega-n\Omega) c^\dagger_q
{b_i^\dagger}^m {b^\dagger}_{i+1}^{n-m} |0\rangle\,,
\end{equation}
where
\begin{equation}
\EqLabel{7} G_0(k,\omega)= {1\over \omega + i\eta - \epsilon_k}
\end{equation}
is the free fermion propagator. With this observation, we find the
 {\em exact} equation of motion for $G(k,\omega)$ to be
\begin{widetext}
\begin{equation}
\nonumber G(k,\omega) = G_0(k,\omega) \left[1-
t_b\sum_{i}^{}{e^{ikR_i}\over \sqrt{N}} \langle 0 | c_k^{}
\hat{G}(\omega) ( c^\dagger_{i-1}+c^\dagger_{i+1})
b_i^\dagger|0\rangle \right]
\end{equation}
because ${\cal V} c_i^\dagger|0\rangle= -t_b (
  c^\dagger_{i-1}+c^\dagger_{i+1}) b_i^\dagger|0\rangle$. This shows
  that the Green's function of interest to us is linked to various
  averages over the Brillouin zone (momentum averages) of the
  $F_1(k,q,\omega)$ Green's functions. To make this more precise, we
  introduce the real-space counterparts of the generalized Green's
  functions defined above, namely:
\begin{equation}
\EqLabel{8} f_n(k,\delta,\omega) ={1\over N} \sum_{q}^{} e^{iq\delta
a} F_n(k,q,\omega) = \sum_{i}^{}{e^{ikR_i}\over \sqrt{N}} \langle 0 |
c_k \hat{G}(\omega) c^\dagger_{i-\delta} {b^\dagger}^n_i|0\rangle\,,
\end{equation}
\begin{equation}
\EqLabel{9} f_{n,m}(k,\delta,\omega) ={1\over N} \sum_{q}^{}
e^{iq\delta a} F_{n,m}(k,q,\omega) = \sum_{i}^{}{e^{ikR_i}\over
\sqrt{N}} \langle 0 | c_k \hat{G}(\omega) c^\dagger_{i-\delta}
{b_i^\dagger}^m {b^\dagger}_{i+1}^{n-m} |0\rangle\,,
\end{equation}
and
\begin{equation}
\EqLabel{10} f_{n,m,p}(k,\delta,\omega) ={1\over N} \sum_{q}^{}
e^{iq\delta a} F_{n,m,p}(k,q,\omega) = \sum_{i}^{}{e^{ikR_i}\over
\sqrt{N}} \langle 0 | c_k \hat{G}(\omega)
c^\dagger_{i-\delta}{b^\dagger}_{i-1}^m {b^\dagger}_{i}^{n-m-p}
{b^\dagger}_{i+1}^{p}|0\rangle\,.
\end{equation}
\end{widetext}
The kets on the right of the resolvent continue to describe a state of
total momentum $k$, however with the fermion fixed at a certain
distance $\delta a$ away from the boson cloud. Here $a$ is the lattice
distance, so that $\delta$ are integers. These definitions do not
necessarily lead to the most ``esthetic'' final equations, but they
are handy and generalize easily to higher dimensions. More symmetric
1D equations can be found if we use the $\sin$ and $\cos$, instead of
complex Fourier transforms.

With these notations, we have the exact equation
\begin{equation}
\EqLabel{11} G(k,\omega) = G_0(k,\omega)
\left[1-t_b\left(f_1(k,1,\omega)+f_1(k,-1,\omega)\right)\right]\,.
\end{equation}

Let us consider now the equation of motion for $F_n(k,q,\omega)$ with
$n\ge1$. When acting on $c^\dagger_q {b_i^\dagger}^n|0\rangle$, both
the boson annihilation and the boson creation part of ${\cal V}$ will
contribute. The boson annihilation contribution can be calculated
exactly, since bosons can only be annihilated at the one site where
they are present. However, the boson creation part can add a boson
either at the site where the cloud is, or to any other site because a
fermion in the state $c^\dagger_q|0\rangle$ is delocalized over the
entire chain. Of course, we do not expect all these outcomes to be
equally likely, and it is this that allows us to make progress.

This is where the MA approximation is made: we only allow new bosons
to be created within at most two-site distance from where the one-site
boson cloud is. In other words, we allow the boson cloud to extend
spatially but not over more than 3 consecutive sites, since we decided
that this is our variational space. After all possible such terms are
accounted for, we find that within this MA approximation:
\begin{widetext}
\begin{align}
  \nonumber F_n&(k,q,\omega)=-t_bG_0(q,\omega-n\Omega) \left[(e^{iqa}
  + e^{-iqa}) nf_{n-1}(k,0,\omega) +[f_{n+1}(k,1,\omega)+
  f_{n+1}(k,-1,\omega)] \right.\\
  &+e^{-iqa+ika}[f_{n+1,1}(k,1,\omega)+ f_{n+1,1}(k,-1,\omega)]
  +e^{iqa}[f_{n+1,n}(k,0,\omega)+f_{n+1,n}(k,-2,\omega)] \nonumber\\
  &+\left. e^{-2iqa+ika}[f_{n+1,1,n}(k,2,\omega)+
  f_{n+1,1,n}(k,0,\omega)] + e^{2iqa-ika}[f_{n+1,n,1}(k,0,\omega)+
  f_{n+1,n,1}(k,-2,\omega)] \right] \,.
\end{align}
\end{widetext}
The terms on the first line come from the exact boson annihilation
part (the first) and the boson creation term where the new boson is
added at the site where the cloud is (the last). The terms in the
second line describe the contributions where the new boson is created
on a NN site of the cloud. The terms on the 3rd line are the
contributions when the new boson is created on a second NN site of the
cloud.

Within the same approximation, we also find
\begin{widetext}
\begin{align}
  \nonumber F_{n,m}&(k,q,\omega)=-t_bG_0(q,\omega-n\Omega)
  \left[(e^{iqa} + e^{-iqa}) mf_{n-1,m-1}(k,0,\omega)
  +(1+e^{2iqa})(n-m)f_{n-1,m}(k,-1,\omega) \right.\\ \nonumber
  &+e^{iqa}[f_{n+1,m}(k,0,\omega)+
  f_{n+1,m}(k,-2,\omega)]+[f_{n+1,m+1}(k,1,\omega)+
  f_{n+1,m+1}(k,-1,\omega)] \\
  &+\left. e^{2iqa-ika}[f_{n+1,m,1}(k,0,\omega)+
  f_{n+1,m,1}(k,-2,\omega)] + e^{-iqa}[f_{n+1,1,n-m}(k,0,\omega)+
  f_{n+1,1,n-m}(k,2,\omega)] \right]
\end{align}
and
\begin{align}
  \nonumber F_{n,m,p}&(k,q,\omega)=-t_bG_0(q,\omega-n\Omega)
  \left[(e^{-2iqa} + 1) mf_{n-1,m-1,p}(k,1,\omega) +(e^{iqa} +
  e^{-iqa}) (n-m-p)f_{n-1,m,p}(k,0,\omega)\right.\\ \nonumber
  &+(1+e^{2iqa})pf_{n-1,m,p-1}(k,-1,\omega)
  +e^{-iqa}[f_{n+1,m+1,p}(k,0,\omega)+ f_{n+1,m+1,p}(k,2,\omega)]\\
  &+\left. [f_{n+1,m,p}(k,1,\omega)+ f_{n+1,m,p}(k,-1,\omega)] +
  e^{iqa}[f_{n+1,m,p+1}(k,0,\omega)+ f_{n+1,m,p+1}(k,-2,\omega)]
  \right] \,.
\end{align}
As before, creation processes are only allowed to add extra bosons so
that the total cloud does not extend over more than 3 consecutive
sites. The annihilation processes are treated exactly, however one has
to be careful when there is a single boson on an outside site of the
cloud. If this is the annihilated boson, the size of the cloud
decreases. As a result:
\end{widetext}
\begin{align}
&f_{n-1,0}(k,\delta,\omega) \rightarrow e^{-ika}
f_{n-1}(k,\delta+1,\omega)\,,\\ &f_{n-1,n-1}(k,\delta,\omega)
\rightarrow f_{n-1}(k,\delta,\omega)\,,\\
&f_{n-1,0,p}(k,\delta,\omega) \rightarrow
f_{n-1,n-1-p}(k,\delta,\omega), \mbox{ if $p< n-1$},\\ &
f_{n-1,0,n-1}(k,\delta,\omega) \rightarrow e^{-ika}
f_{n-1}(k,\delta+1,\omega)\,,\\ & f_{n-1,m,0}(k,\delta,\omega)
\rightarrow e^{ika} f_{n-1,m}(k,\delta-1,\omega), \mbox{ if $m<
n-1$},\\ & f_{n-1,n-1,0}(k,\delta,\omega) \rightarrow e^{ika}
f_{n-1}(k,\delta-1,\omega)\,.
\end{align}
 All these identities follow directly from the definitions of
Eqs.~(\ref{8})-(\ref{10}).

We have thus generated an infinite system of coupled equations of
motion linking various Green's functions with a total of $n$ bosons to
Green's functions with $n-1$ and $n+1$ bosons. The only approximation
is the restriction on the size of the allowed boson cloud. A solution
of this system will give $G(k,\omega)$ within this MA approximation,
together with all the other generalized Green's functions from which
we can extract additional information on the structure of the polaron
cloud.\cite{MAno}

This solution is straightforward to obtain in terms of the momentum
averaged Green's functions $f_n$. First, note that only a finite
number of these are needed, for a given value of $n$, in order to be
able to calculate everything else. For example, only
$f_{n,m,p}(k,\delta,\omega)$ with $|\delta|\le 2$ appear on the right
hand side of all these equations, and similar bounds can be found for
the other functions. We therefore first generate a set of recurrence
equations for these quantities, using Eqs.~(\ref{8})-(\ref{10}). These
read:
\begin{widetext}
\begin{align}
 \nonumber f_n&(\delta)=-t_b\left[g_0(\delta+1,\omega_n) +
  g_0(\delta-1,\omega_n) \right] nf_{n-1}(0) -t_b g_0(\delta,\omega_n)
  [f_{n+1}(1)+ f_{n+1}(-1)] \\ \nonumber
  &-t_bg_0(\delta-1,\omega_n)e^{ika}[f_{n+1,1}(1)+ f_{n+1,1}(-1)] -t_b
  g_0(\delta+1,\omega_n) [f_{n+1,n}(0)+ f_{n+1,n}(-2)] \\
  &-t_bg_0(\delta-2,\omega_n) e^{+ika}[f_{n+1,1,n}(2)+ f_{n+1,1,n}(0)]
  - t_bg_0(\delta+2,\omega_n)e^{-ika}[f_{n+1,n,1}(0)+
  f_{n+1,n,1}(-2)]\,,\EqLabel{11b}
\end{align}

\begin{align}
\nonumber f_{n,m}&(\delta)=-t_b[g_0(\delta+1,\omega_n)
  +g_0(\delta-1,\omega_n)] mf_{n-1,m-1}(0) -t_b[g_0(\delta,\omega_n)
  +g_0(\delta+2,\omega_n)] (n-m)f_{n-1,m}(-1) \\ \nonumber
  &-t_bg_0(\delta+1,\omega_n)[f_{n+1,m}(0)+ f_{n+1,m}(-2)]-
  t_bg_0(\delta,\omega_n)[f_{n+1,m+1}(1)+ f_{n+1,m+1}(-1)] \\
  \EqLabel{12} & -t_bg_0(\delta+2,\omega_n)e^{-ika}[f_{n+1,m,1}(0)+
  f_{n+1,1,n}(-2)]-t_bg_0(\delta-1,\omega_n)[f_{n+1,1,n-m}(0)+
  f_{n+1,1,n-m}(2)]\,,
\end{align}
and
\begin{align}
\nonumber f_{n,m,p}&(\delta)=-t_b[g_0(\delta-2,\omega_n)
  +g_0(\delta,\omega_n)] mf_{n-1,m-1,p}(1) -t_b[g_0(\delta-1,\omega_n)
  +g_0(\delta+1,\omega_n)] (n-m-p)f_{n-1,m,p}(0)\\ \nonumber
  &-t_b[g_0(\delta,\omega_n) +g_0(\delta+2,\omega_n)]
  pf_{n-1,m,p-1}(-1) -t_bg_0(\delta-1,\omega_n)[f_{n+1,m+1,p}(0)+
  f_{n+1,m+1,p}(2)]\\ \EqLabel{13} &
  -t_bg_0(\delta,\omega_n)[f_{n+1,m,p}(1)+ f_{n+1,m,p}(-1)]
  -t_bg_0(\delta+1,\omega_n)[f_{n+1,m,p+1}(0)+ f_{n+1,m,p+1}(-2)]\,.
\end{align}
\end{widetext}
Here we used the shorthand notations $f_n(\delta)\equiv
f_n(k,\delta,\omega)$ etc. and $\omega_n\equiv \omega-n\Omega$ in
order to simplify the notation. Also,
\begin{equation}
\EqLabel{14} g_0(\delta,\omega) = {1\over N} \sum_{k}^{} e^{ik\delta
a} G_0(k,\omega)
\end{equation}
are the free propagators in real space, which can be calculated
analytically.\cite{gre,MA0} For any given number $n$ of bosons, the
needed Green's 
functions are $f_{n,m,p}(k,\delta,\omega)$ with $|\delta|\le 2$,
$f_{n,m}(k,\delta,\omega)$ with $|\delta|\le 3$ and
$f_n(k,\delta,\omega)$ with $-3\le \delta\le 4$. Once we know these,
we can calculate all other $f$ and $F$ generalized Green's functions.

To solve this set of recurrence equations, we order all Green's
functions with a given $n$ in a vector $V_n$ of dimension
$7n+1+5n(n-1)/2$. Then, for any $n\ge1$, Eqs.~(\ref{11})-(\ref{13})
map into the matrix recurrence equations
\begin{equation}
\EqLabel{15} V_n= \alpha_n(k,\omega) V_{n-1} + \beta_n(k,\omega)
V_{n+1}\,,
\end{equation}
where $\alpha_n(k,\omega)$ and $\beta_n(k,\omega)$ are sparse matrices
which can be read directly from Eqs.~(\ref{11})-(\ref{13}). The
solution, for any $n\ge 1$, has the general form\cite{MA0,MA1}
\begin{equation}
\EqLabel{17} V_n = A_n(k,\omega) V_{n-1}\,,
\end{equation}
 where $A_n(k,\omega)$ are given by the continued fractions
\begin{equation}
\EqLabel{18} A_n(k,\omega) =
\frac{\alpha_n(k,\omega)}{1-\beta_n(k,\omega) A_{n+1}(k,\omega)}
\end{equation}
starting from a large value $N$ with $A_N(k,\omega)=0$. The physical
motivation for this choice has been discussed at length elsewhere.\cite{MA0}
Briefly, it is because clouds with too many bosons
$N\rightarrow\infty$ are too expensive energetically, and therefore
very unlikely to be observed. Hence, the propagators into such states
must become vanishingly small for a large enough $N$.  In
practice one increases $N$ until the matrices $A_n(k,\omega)$ are
converged to within any accuracy one chooses.
 
Consider now Eq.~(\ref{17}) for $n=1$. The entries in $V_0$ are the
various $f_0(k,\delta,\omega)=e^{ik \delta a} G(k,\omega)$ (see
Eq.~(\ref{8}); the functions $f_{n,m}$ and $f_{n,m,p}$ are defined
only for $n\ge 2$). As a result, once we know the matrix
$A_1(k,\omega)$, we find
\begin{equation}
f_1(k,\pm 1,\omega) = a^{\pm}(k,\omega) G(k,\omega)\,,
\end{equation}
where $a^{\pm}(k,\omega)$ are combinations of the appropriate matrix
elements of $A_1(k,\omega)$ and $e^{ik \delta a} $ factors. Using this
in Eq.~(\ref{11}) gives us the standard solution
\begin{equation}
\EqLabel{19} G(k,\omega) = {1\over \omega +i \eta - \epsilon_k
-\Sigma(k,\omega)}
\end{equation}
with a self-energy, at this level of MA approximation, of
\begin{equation}
\EqLabel{20} \Sigma(k,\omega) = -t_b\left[a^{+}(k,\omega)+
a^{-}(k,\omega)\right]\,.
\end{equation}

In some limiting cases the model has certain symmetries that will
insure that many of the generalized Green's functions vanish
identically. A simple example is the case $t_f=0$.~\cite{AEF10} Then
the free fermion dispersion vanishes, $\epsilon_k =-2t_f
\cos(ka)\rightarrow0$, and so:
\begin{equation}
g_0(\delta,\omega) \rightarrow \delta_{\delta,0} {1\over \omega
  +i\eta}\;.
\end{equation}
This simplifies Eqs.~(\ref{11})-(\ref{13}) significantly and one can
show that many of the $f$ functions are identically zero. For example,
only states $f_{2n,n}$ and $f_{2n+1,n},f_{2n+1,n+1}$ survive. This
becomes obvious if one considers what kinds of boson clouds can be
formed if one acts with ${\cal V}$ repeatedly on any one-fermion
state. By continuity, one expects that for small $t_f$, these
functions will still be very small although not precisely zero. This
fact could be used to speed up the calculation in this limit, by
removing these kets from the variational set and thus decreasing the
size of all these matrices.

In fact, it is known that $g_0(\delta,\omega)$ decreases exponentially
with increasing distance $|\delta|$ for values $\omega < - 2d t_f$,
i.e. below the free fermion continuum ($d$ is the
dimension).\cite{gre}  If we
are interested in low-energy properties at $\omega < - 2d t_f$, and
especially since $\omega_n = \omega-n\Omega$ appears in
Eqs.~(\ref{11})-(\ref{13}), we see that most of the terms on the right
are multiplied by exponentially small functions if $|\delta|>0$ in
their corresponding $g_0(\delta,\omega_n)$ factor. This explains the
earlier statement that the approximation ignores only exponentially
small contributions. One can check that going to more extended clouds
will bring in more terms in Eqs.~(\ref{11})-(\ref{13}), but their
$g_0(\delta,\omega_n)$ factors will be even smaller. This argument
only becomes problematic when the ground-state energy is not too far
below the free-fermion continuum and $\Omega/t_f \rightarrow 0$.  In
this case, $g_0(\delta,\omega_n)$ still decays exponentially with
increasing $\delta$ but very slowly, and much more extended clouds may
form with significant probability. For the Holstein model, this leads
to quantitative discrepancies in the MA prediction for intermediate
couplings, where the polaron cloud can extend over many
sites.~\cite{WF98a,AFT10} At strong couplings,  the Lang-Firsov
approach~\cite{LF62r} gives a small, one-site cloud polaron, and MA
with a one-site cloud restriction becomes asymptotically exact. No
exact asymptotic solution is known for the Edwards model, therefore
there is no guarantee that in the limit $\Omega/t_f \rightarrow 0$, a
3-site cloud will provide a good description.  However, as we show
below, MA allows us to decide, with a high level of confidence,
whether it provides a reasonable description.

Finally, let us comment on why we chose to allow a three-site cloud,
as oppose to a one-site or two-site one. Consider what would happen if
we restricted the variational space to single-site clouds only. Then,
we can set all functions $f_{n,m}$ and $f_{n,m,p}$ identically to
zero. From Eq.~(\ref{11}), we see that the recurrence relation that
now links together only the $f_n(\delta)$ with $\delta=-1,0,1$ does
not contain any dependence of $k$ in any of its terms, resulting in a
self-energy that is independent of $k$. For example, for $t_f=0$ this
would mean that the polaron is also dispersionless. We know that this
is not the case due to three-site boson processes,~\cite{AEF07}
therefore we need to include at least such cloud structures in the
calculation to capture this. If $\Omega$ is not too small, one can
argue that much more extended clouds are unlikely for energetical
reasons: bosons far from the particle cost energy $\Omega$ to create
yet are unlikely to interact with the particle because of the
separation. Of course, one can increase the size of the cloud
systematically until convergence is achieved. Such an exercise allows
one to uncover the physics essential for explaining the properties of
the polaron, from the importance of various terms play.

\subsection{MA results for the 1D Edwards model}

We can analyze the accuracy of the MA approximation for the Edwards
model in 1D, where accurate numerical results have been obtained using
variational Exact Diagonalization (ED).~\cite{AEF07} We begin with the
correlated transport regime $t_f/t_b \ll 1$ (cf. Fig.~1 in
Ref.~\onlinecite{AEF07}), where polaron motion becomes possible only
through emission/absorption of bosons. This will be the regime of main
interest to us when discussing the 2D problem.

\begin{figure}[t]
\includegraphics[width=\columnwidth]{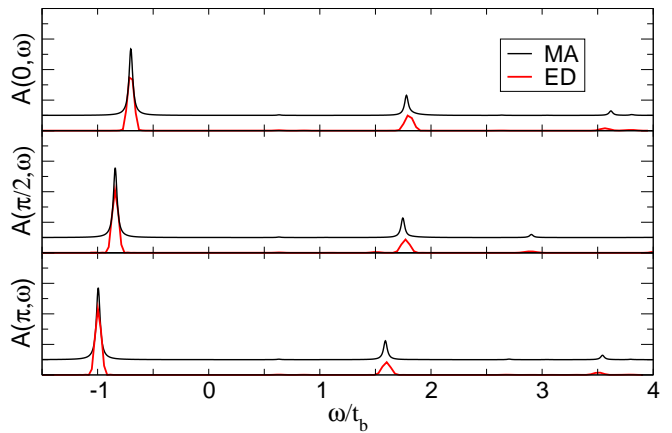}
\caption{(Color online) Spectral weights $A(k,\omega)$ for $k=0$
  (top), $k={\pi\over 2}$ (middle) and $k=\pi$ (bottom) from MA (black
  thin lines) and ED (red thick lines). The MA results are shifted
  upwards to ease the comparison. Parameters are $\Omega/t_b=2,
  t_f/t_b=0.1, \eta/t_b=0.02$.
\label{fig1}}
\end{figure}

\begin{figure}[t]
\includegraphics[width=\columnwidth]{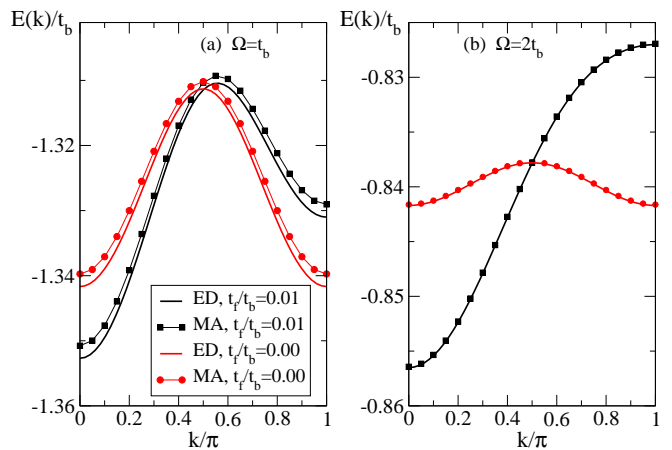}
\caption{(Color online) Polaron energy $E(k)$ in units of $t_b$
  vs. $k$, for the parameters indicated. Lines show ED results,
  symbols show  MA results (dashed lines are guides to the eye).
\label{fig2}}
\end{figure}

In Fig.~\ref{fig1} we compare spectral weights $A(k,\omega)$ at
$k=0,{\pi\over 2}$ and $\pi$ for $\Omega/t_b=2, t_f/t_b=0.1$. The
agreement between MA (thin line, shifted upwards) and ED (thick line)
results is very satisfactory, especially for the features with larger
weights. The agreement is of similar quality throughout the whole
Brillouin zone (not shown). The most prominent features is the
low-energy polaron band, which disperses very little (its bandwidth is
comparable to the smallest energy scale in the problem, $t_f$), and
has a  {\em qp} weight that varies little with $k$.

\begin{figure}[b]
\includegraphics[width=\columnwidth]{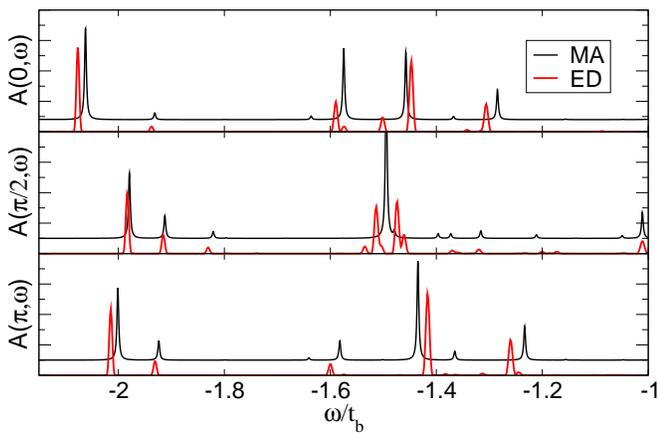}
\caption{(Color online) Spectral weights $A(k,\omega)$ for $k=0$
  (top), $k={\pi\over 2}$ (middle) and $k=\pi$ (bottom) from MA and
  ED. The MA results (black thin lines) are shifted upwards. The
  parameters are $\Omega/t_b=0.5, t_f/t_b=0.04, 
  \eta/t_b=0.002$.
\label{fig3}}
\end{figure}

For a more detailed comparison, we focus on the polaron band, and plot
its energy $E(k)$ vs $k$ in Fig.~\ref{fig2}. As expected, for $t_f=0$
the dispersion corresponds to pure 2nd NN hopping $-2t_2\cos(2ka)$,
where $t_2$ is dynamically generated through the 3-site, 3-boson
processes. For $t_f\ne0$, an additional term $-2t_f^*\cos(ka)$ with a
renormalized transfer amplitude $t_f^*$ is also
present. Figs.~\ref{fig2}a,b show quite good agreement between MA and
ED. As expected, the agreement is better for the larger $\Omega/t_b=2$
value, where the probability of more extended clouds is
reduced. However, even for the smaller $\Omega/t_b=1$, MA captures the
dispersion quite accurately: most of the difference to the ED results
is an overall shift independent of $k$. This suggests that more
extended clouds will not further renormalize the effective hopping
integrals, only lower the overall polaron formation energy. This is
reasonable, because longer range effective hopping terms are more
complicated to generate and involve many more sites than the 3-site,
3-boson process responsible for $t_2$. It is therefore probably safe
to conclude that MA accurately describes the polaron's dynamics in the
$t_f/t_b \ll 1 $ region, so long as $\Omega/t_f$ is not too small so
that the spatial Green's functions decay exponentially reasonably
fast.

\begin{figure}[t]
\includegraphics[width=\columnwidth]{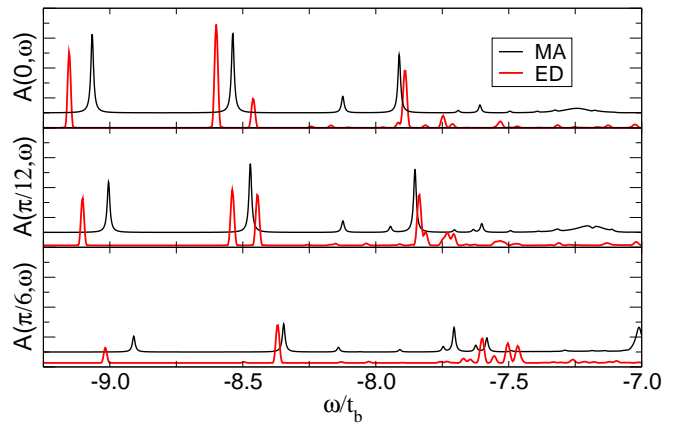}
\caption{(Color online) Spectral weights $A(k,\omega)$ for $k=0$
  (top), $k={\pi\over 12}$ (middle) and $k=\pi/6$ (bottom) from MA and
  ED. Again the MA results are shifted upwards. The
  parameters are $\Omega/t_b=0.5, t_f/t_b=4, \eta/t_b=0.006$.
\label{fig4}}
\end{figure}

As $\Omega/t_b$ is decreased we expect to move in the regime of
``strong fluctuations''.~\cite{AEF07} The low-energy part of the
spectral weight in this regime is shown in Fig.~\ref{fig3} for
$\Omega/t_b=0.5, t_f/t_b=0.04$. Because the cost of exciting bosons is
reduced, we expect to see many more lower-energy features than in
Fig.~\ref{fig1}, and this is indeed the case. MA shows reasonable
agreement with ED at the lower energies, however, some of the
higher-energy peaks are either missing, or displaced, or combined in a
single feature. In order to properly describe these peaks, it is
likely that one would need to go to the MA$^{(1)}$ or higher level
approximations, where bosons are allowed to exist far from the main
polaron cloud (cf. Fig.~\ref{fig0b}(b)).  Such states are essential in
describing the polaron+one boson continuum starting at
$E_{gs}+\Omega$,\cite{num1,MA1} and indeed, it is at these energies that the
disagreements between MA and ED become more visible.  If, however, the
focus is on understanding the polaron band and if the polaron
bandwidth is less than $\Omega$ (as is the case we discuss in 2D,
below), then inclusion of these states is not absolutely necessary:
doing so will improve the quantitative agreement, of course, but will
not change qualitatively the polaron dispersion.

Next we explore the ``incoherent'' or ``diffuse'' region of the
parameter space, where $t_b/\Omega \gg 1$ while $t_b/t_f \ll
1$.~\cite{AEF07} Since this implies $\Omega \ll t_f$, this is where
the MA approximation is expected to be least accurate. In
Fig.~\ref{fig4} we show comparisons of the spectral function
$A(k,\omega)$ vs $\omega$ for small values of $k$. At $k=0$ (upper
panel), ED shows two sharp peaks, associated with the polaron and the
second bound state,\cite{num1,MA1} followed by the polaron+one boson
continuum at 
$E_{gs}+\Omega$, and then more features at higher energies. MA finds
the two peaks associated with the bound polaron states, shifted to
slightly higher energies, but the continuum is absent since, as
discussed, it is not included in the variational space at this MA
level. As $k$ increases, the spectral weight (equal to the area under
the peak) in the polaron band decreases extremely fast, similar to
what happens for Holstein polarons at weak
coupling,~\cite{St96,MA1,WF97,FT07} and most of the weight shifts to
roughly $\omega=\epsilon_k$. The lower panels of Fig.~\ref{fig4}
illustrate this fast decrease in the polaron band spectral weight as
$k$ increases.~\cite{WF97} We also see that even though the polaron
band predicted by MA is shifted upwards, this shift is again not
strongly $k$ dependent. The {\em qp} weight is also in good agreement
with the ED results, so MA is still doing a reasonable job in
predicting the {\em qp} weight and effective mass. However, at larger
$k$, the absence of the continuum within this level of MA
approximation is expected to lead to an overestimate of the polaron
bandwidth.\cite{MA0}

\begin{figure}[t]
\includegraphics[width=\columnwidth]{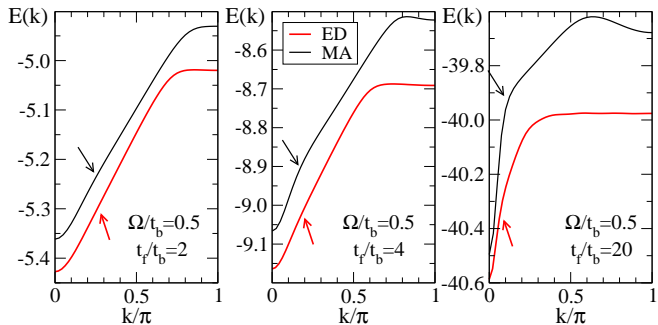}
\caption{(Color online) Polaron energy $E(k)$ in units of $t_b$
  vs. $k$, for the parameters indicated. Dashed lines show MA
  results, full lines correspond to ED results.  The arrows
  roughly mark a discontinuity in the slope of the dispersion.
\label{fig5}}
\end{figure}

This is indeed the case, as shown in Fig.~\ref{fig5}, where we compare
the polaron dispersions $E(k)$ vs. $k$ for $\Omega=0.5 t_b$ and
$t_f/t_b =2,4,20$, so that $\Omega/2t_f = 0.125, 0.0625 $ respectively
$0.0125$. The agreement at large $k$ worsens as $\Omega/t_f\rightarrow
0$. This is because in all these cases the bandwidth of the MA $E(k)$
exceeds $\Omega$, which is unphysical: the polaron band must always
stay below the continuum. It follows that in this regime, the polaron
+ one boson continuum plays the key role in defining the polaron
bandwidth. To describe it within MA, one needs to allow at least one
boson to be far away from the polaron, i.e. to go to the MA$^{(1)}$
level. The simpler MA$^{(0)}$ level approximation used here is
certainly untrustworthy at large $k$ in such cases. However, it still
provides reasonably good description near $k=0$, as shown in
Figs.~\ref{fig4} and \ref{fig5}. For example, the ED polaron
dispersion shows a ``kink'' at small $k$, whose origin we do currently
understand. This feature is more pronounced as $\Omega/t_f$ decreases,
and is indicated by the arrows in Fig.~\ref{fig5}.  MA also replicates
it, albeit at somewhat different location. Below this value, the two
dispersions seem to match quite well (up to an overall shift). We
conclude that even in this least favorable regime and for this lowest
level of MA, we can still use it to get reasonable estimates for the
polaron effective mass and ground-state {\em qp} weight. This is
verified by the data presented in Table \ref{tb1}, where we compare
the effective masses (in dimensionless units),
\begin{equation}
\frac{1}{m^*}=\frac{1}{a^2t_b} \left.\frac{\partial^2E(k)}{ \partial
  k^2}\right|_{k=0}\,,
\end{equation} 
calculated with the two methods.

To summarize, these comparisons confirm that if the MA polaron
bandwidth is less than $\Omega$ so that the continuum can be ignored,
then this level of MA approximation suffices to describe $E(k)$ with
good accuracy throughout the Brillouin zone. This is generically
expected to hold so long as $\Omega/t_f$ is not too small, in other
words for most of the parameter space. In the limit of small
$\Omega/t_f$, the lowest MA approximation suffices for a reasonable
description of GS properties, but one needs to go to higher MA levels
and enlarge the variational space if one wants to have a good
description of $E(k)$ in all the Brillouin zone. The same holds true
if one wants an accurate description at energies above the polaron
band. Because the MA approximations generically satisfy with good
accuracy spectral weight sum rules,\cite{MA0,MA1} one expects even the
lowest MA 
level to identify quite correctly where significant spectral weight
appears in the spectrum. This is indeed the case, as shown in the
comparisons provided here (regarding Fig.~\ref{fig5}, one must
remember that there is essentially no spectral weight in the {\em qp}
band once it gets close to the continuum). However, in order to
capture finer details, one needs to work harder by suitably enlarging
the variational space. In practice, this requires one to figure out
the corresponding equivalent of Eqs.~(\ref{11})-(\ref{13}) and find an
efficient way to solve them.

In the following, we focus on the polaron dispersion of the 2D Edwards
model in the limit $t_f/ t_b \rightarrow 0$ for a finite $\Omega/t_b$
ratio. In this case, the MA level introduced here---suitably
generalized to 2D---is sufficient for an accurate description of
$E(k)$ in the entire Brillouin zone, therefore we do not need to go to
a higher level.

\begin{table}[t]
  \begin{tabular}{ |c | c | c | c|}\hline
    $\qquad$ $\Omega$ $\qquad$ & $\qquad$ $t_f$ $\qquad$ & $\qquad$
    $m^*_{ED}$ $\qquad$ & $\qquad$ $m^*_{MA}$ $\qquad$\\ \hline\hline
    1.0 & 0.00 & 17.71 & 18.12\\ \hline 1.0 & 0.01 & 15.75 & 15.51\\
    \hline 2.0 & 0.00 & 129.28 & 129.98\\ \hline 2.0 & 0.01 & 44.68 &
    44.77\\ \hline 0.5 & 2.00 & 1.54 & 1.39\\ \hline 0.5 & 4.00 & 0.61
    & 0.47\\ \hline 0.5 & 20.00 & 0.052 & 0.041\\ \hline
  \end{tabular}
\caption{Comparison of effective masses as predicted by
  ED~\cite{Alpr10} and MA, for several values of $\Omega$ and $t_f$
  where $t_b=1$.}
\label{tb1}
\end{table}

\subsection{MA for the 2D square lattice}

To generate the MA equations---within the 3-site cloud variational
space---for a 2D square lattice, we follow the steps outlined in the
previous section. The only complication is that now the 2-site clouds
can be aligned either along the $x$ or $y$ directions, so we need two
types of 2-site cloud generalized Green's functions,
\begin{equation}
\EqLabel{19b} F^{(\ev)}_{n,m}(\kv,\qv,\omega) = \sum_{i}^{}
e^{i(\kv-\qv)\Rv_i}\langle 0| c_\kv \hat{G}(\omega) c^\dagger_\qv
{b_i^\dagger}^m {b^\dagger}_{i+\ev}^{n-m}|0\rangle
\end{equation}
with the associated
\begin{align}
\EqLabel{20b} &f^{(\ev)}_{n,m}(\kv,\dv,\omega) = {1\over N}
\sum_{q}^{} e^{i\qv\dv a} F^{(\ev)}_{n,m}(\kv,\qv,\omega)
\nonumber\\&\;\;= \sum_{i}^{}{e^{i\kv\Rv_i}\over \sqrt{N}} \langle 0 |
c_\kv \hat{G}(\omega) c^\dagger_{i-\dv} {b^\dagger}_{i}^{m}
{b^\dagger}_{i+\ev}^{n-m}|0\rangle\,,
\end{align}
where $\ev=(1,0)=\x$ or $\ev=(0,1)=\y$, each site index is
two-dimensional, e.g.  $i=(i_x,i_y)$, and we used the shorthand
notation $i+\ev = (i_x+1,i_y)$ if $\ev=\x$ etc. As in 1D, we ask that
$1\le m\le n-1$ to avoid overlap with the 1-site cloud functions.

For the 3-site clouds, we define:
\begin{widetext}
\begin{equation}
\EqLabel{21} F^{(\ev,\ev')}_{n,m,p}(\kv,\qv,\omega) = \sum_{i}^{}
e^{i(\kv-\qv)\Rv_i}\langle 0| c_\kv \hat{G}(\omega) c^\dagger_\qv
{b^\dagger}_{i-\ev}^m {b_i^\dagger}^{n-m-p}
{b^\dagger}_{i+\ev'}^{p}|0\rangle \,,
\end{equation}
\begin{equation}
\EqLabel{22} f^{(\ev,\ev')}_{n,m,p}(\kv,\dv,\omega) = {1\over N}
\sum_{q}^{} e^{i\qv\dv a} F^{(\ev,\ev')}_{n,m,p}(\kv,\qv,\omega) =
\sum_{i}^{}{e^{i\kv\Rv_i}\over \sqrt{N}} \langle 0 | c_\kv
\hat{G}(\omega) c^\dagger_{i-\dv}{b^\dagger}_{i-\ev}^m
{b_i^\dagger}^{n-m-p} {b^\dagger}_{i+\ev'}^{p} |0\rangle\,.
\end{equation}
\end{widetext}

\begin{figure}[t]
\includegraphics[width=0.5\columnwidth]{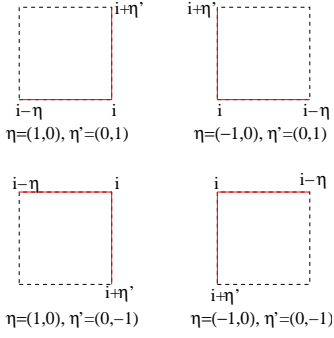}
\caption{(Color online) Sketch of our choice of $\ev, \ev'$ indices
  for the $n$-boson non-collinear 3-site boson clouds on the 2D square
  lattice. In all cases, $m\ge 1$ bosons are at site $i-\ev$, $p\ge 1$
  bosons are at site $i+\ev'$ and the remaining $n-m-p$ bosons are at
  site {i}.
\label{fig6}}
\end{figure}

To describe collinear clouds we take $\ev =\ev'$, and we again need
only keep $\ev=\x,\y$ for the two possible orientations. For
non-collinear clouds we have 4 different distinct possibilities which
we choose as $\ev=\pm \x =(\pm 1, 0)$ and $\ev' =\pm \y=(0, \pm 1)$,
as sketched in Fig.~\ref{fig6}. They are inequivalent except when
$m+p=n$, so that there are bosons only on the opposite diagonal
sites. Again, only $m\ge1, p\ge1$ are allowed.

The equations for the various $F$ and $f$ functions are generated just
as in the 1D case, using the restriction that boson addition
contributions cannot extend the boson cloud to more than 3 neighboring
sites. This again leads to a set of equations for a finite number of
$n$-boson functions $f$ which depend only on $n-1$ and $n+1$ boson $f$
functions. While straightforward to obtain, these equations are very
lengthy (e.g. the equation for $f_n(\kv,\dv,\omega)$ has 64
contributions on the right-hand side), and we do not list all of them
here. As an example, the relevant equations for the 3-site cloud
functions are listed below, using again the shorthand notations $
f^{(\ev,\ev')}_{n,m,p}(\dv) \equiv
f^{(\ev,\ev')}_{n,m,p}(\kv,\dv,\omega)$ and $\omega_n=\omega-\Omega$:
\begin{widetext}
\begin{eqnarray}
\nonumber f^{(\ev,\ev')}_{n,m,p}(\dv) =&&-mt_b \sum_{\dv'=\pm \x,
  \pm\y}^{}g_0(\dv+\dv' -\ev,\omega_n) f^{(\ev,\ev')}_{n-1,m-1,p}(\ev)
  -(n-m-p)t_b \sum_{\dv'=\pm \x, \pm\y}^{}g_0(\dv+\dv',\omega_n)
  f^{(\ev,\ev')}_{n-1,m,p}(0) \\ \nonumber && -pt_b \sum_{\dv'=\pm \x,
  \pm\y}^{}g_0(\dv+\dv' +\ev',\omega_n)
  f^{(\ev,\ev')}_{n-1,m,p-1}(-\ev') -t_b g_0(\dv+\ev,\omega_n)
  \sum_{\dv'=\pm \x, \pm\y}^{} f^{(\ev,\ev')}_{n+1,m+1,p}(\ev+\dv') \\
  \EqLabel{25} && -t_b g_0(\dv,\omega_n) \sum_{\dv'=\pm \x, \pm\y}^{}
  f^{(\ev,\ev')}_{n+1,m,p}(\dv') -t_b g_0(\dv-\ev',\omega_n)
  \sum_{\dv'=\pm \x, \pm\y}^{}
  f^{(\ev,\ev')}_{n+1,m,p+1}(\dv'-\ev')\,.
\end{eqnarray}
\end{widetext}

\begin{figure}[b]
\includegraphics[width=0.7\columnwidth]{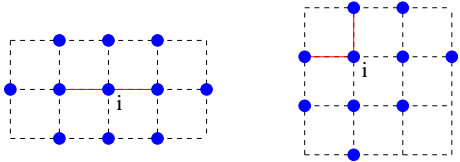}
\caption{(Color online) Pictorial description of the minimal sets of
  values $\dv$ needed for a collinear (left) and non-collinear (right)
  3-site cloud generalized function $f_{n,m,p}(\dv)$. Red solid lines
  indicate the boson cloud centered at $i$, while the blue dots mark
  the possible locations of the fermion.  The distances from the
  fermion to the site $i$ define the needed set of $\dv$ values.
\label{fig7}}
\end{figure}

An analysis of the terms occurring on the right-hand sides of these
equations allows one to identify the minimum set of values $\dv$
needed for the various $f$ functions. These are shown explicitly for a
collinear and one non-collinear cloud in Fig.~\ref{fig7}. There, the
thick lines mark the position of the cloud, which is centered at $i$,
while the dots show the locations of the fermion. The distances from
the fermion to the site $i$ define the needed set of $\dv$ values for
these clouds. The other 3-site clouds have sets related by appropriate
symmetries. The sets for the 2-site and 1-site clouds are found
similarly.

Once this is done, the solution follows that used in 1D. All functions
with a given $n$ are collected in a vector $V_n$. The equations of
motions again reduce to matrix recurrence equations
$V_n=\alpha_n(k,\omega) V_{n-1} + \beta_n(k,\omega) V_{n+1}$ which are
solved in precisely the same way. Of course, the dimension of $V_n$ is
now significantly increased, $\dim(V_n)= 31n^2-15n-3$, and therefore
the various matrices $A_n, \alpha_n, \beta_n$ needed are larger than
in 1D, however they are still very manageable. Most results shown
below converged with relative errors less than $10^{-4}$ if we started
from $A_N=0$ with $N=9$ or less, so that the largest vectors'
dimension is below 2000. Moreover, their dimensions decrease fast as
$n$ decreases, so the solution is still very efficient. In the plots
shown below, a data point typically takes around a minute or less to
generate.

We begin with a thorough analysis of the most interesting case, when
$t_f=0$. In this case, only the 3-site, 3-boson terms already
discussed will lead to dynamic generation of a polaron dispersion. We
already know from the 1D case that we expect the generation of terms
of the type $\sim -2t_3[\cos(2k_xa) + \cos(2k_ya)]$ from the collinear
clouds (on the 2D lattice, this corresponds to effective 3rd NN
hopping, hence $t_3$). However, because of the closed path (Trugman
loops) processes that are now also possible, we also expect dynamic
generation of second NN hopping, leading to terms of the type $\sim
-2t_2[\cos((k_x+k_y)a) + \cos((k_x-k_y)a)]$. Altogether, then, in this
case the polaron dispersion should be well described by
\begin{eqnarray}
\nonumber E(\kv) &=& E_P -2t_2[\cos((k_x+k_y)a) +
  \cos((k_x-k_y)a)-2]\\ \EqLabel{26} &&- 2t_3[\cos(2k_xa) +
  \cos(2k_ya)-2]\,,
\end{eqnarray}
where $E_P=E(\kv=0)$ is the polaron ground state energy. From the 1D
analysis, we expect that $t_2$ and $t_3$ should be quite accurately
predicted by MA, while $E_P$ is only accurate at fairly high $\Omega$
and becomes systematically underestimated as $\Omega$ decreases.

\begin{figure}[t]
\includegraphics[width=\columnwidth]{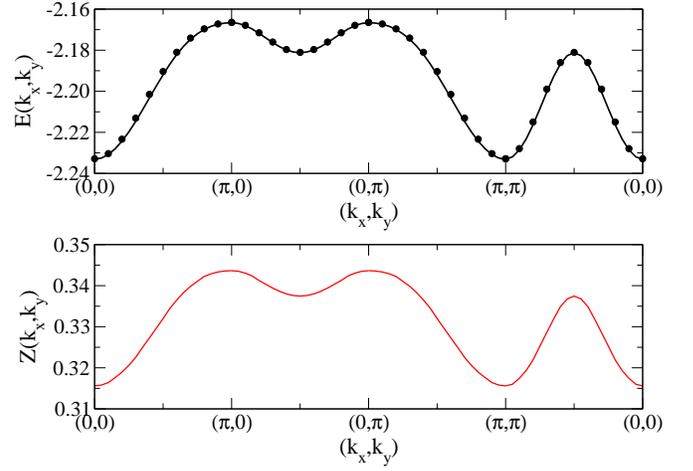}
\caption{(Color online) Polaron dispersion $E(\kv)$ (upper panel) and
  {\em qp} weight $Z(\kv)$ (lower panel) along the high-symmetry
  directions in the Brillouin zone of the 2D square lattice, for
  $t_f=0, t_b=\Omega=1$. Lines show MA results, the symbols are fits
  to Eq.~(\ref{26}).
\label{fig8}}
\end{figure}

In Fig.~\ref{fig8} we plot the polaron dispersion $E(\kv)$ and {\em
qp} weight $Z(\kv)$ along lines of high symmetry in the Brillouin
zone. Interestingly, the two curves have similar profiles, however the
{\em qp} weight changes very little in real terms. This is somewhat
reminiscent of Holstein polarons in the strong-coupling limit, which
also have an almost constant {\em qp} weight throughout the Brillouin
zone. However, in that limit their {\em qp} weight is exponentially
small and the effective mass is exponentially large, whereas here the
{\em qp} weight is still considerable, as is the polaron
bandwidth. This shows that very different physics gives rise to this
behavior. Indeed, the small Holstein polaron has a small-size cloud
with a large average number of bosons. This explains its very large
effective mass (due to vanishing overlap of the spatially small
polaron cloud), the very small {\em qp} weight (free fermion
contribution to the wavefunction is very small), and its weak
sensitivity to $\kv$ (states that are nearly localized in real space
are ``extended'' in $\kv$-space). In contrast, for the Edwards model
polaron at $t_f=0$, all the dispersion is due to the existence of
bosons through the boson-assisted hopping. As illustrated in
Fig.~\ref{fig0}, the free fermion state mixes with the various
fermion+bosons states to give rise to the effective 2nd and 3rd NN
hopping, so the fairly significant {\em qp} weight throughout the
Brillouin zone is not surprising. It is worth emphasizing again that
the doubling of the Brillouin zone is due to the boson-modulated
hopping. In fact, the resulting dispersion is somewhat reminiscent of
that of a doped hole in a cuprate layer, although there the minimum is
at $({\pi\over 2a}, {\pi\over 2a})$, which here is a saddle point.

The symbols in Fig.~\ref{fig8} are fits to
Eq.~(\ref{26}). Specifically, we used the MA values for $E(0,0),
E({\pi\over2},{\pi\over2})$ and $E(0,\pi)$ in order to extract $E_P,
t_2$ and $t_3$ from Eq.~(\ref{26}), and use these to generate the
dispersion-fit in the entire Brillouin zone. The agreement is very
reasonable, backing up our assumptions about polaronic physics in this
regime. The only problem is that $t_3\ll t_2, E_P$, and as a result
the level of confidence in extracting this parameter is not very
high. For example, if we use $E(0,{\pi\over2})$ instead of
$E({\pi\over2},{\pi\over2})$ as the 3rd point, the value of $t_3$
changes from $0.0023$ to $0.0018$ ($E_P, t_2$ remain unchanged) but
the agreement between the overall fit and $E(\kv)$ is visibly poorer.

\begin{figure}[t]
\includegraphics[width=\columnwidth]{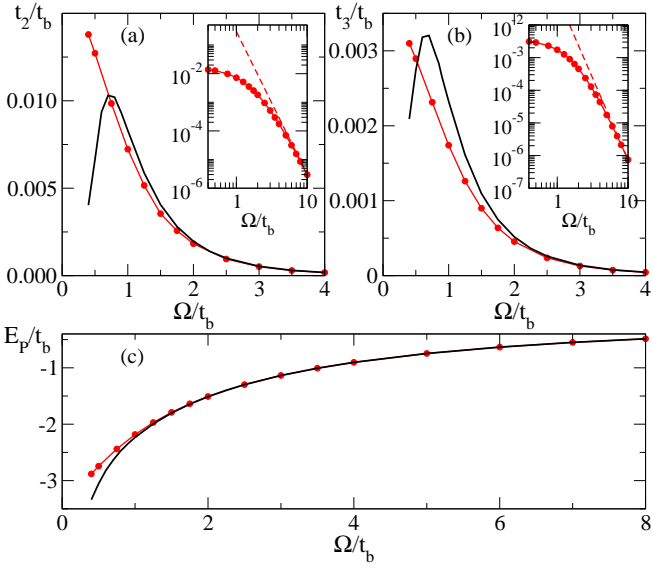}
\caption{(Color online) Effective $t_2/t_b$ in (a), $t_3/t_b$ in (b)
  and $E_P/t_b$ in (c), vs. $\Omega/t_b$ for $t_f=0$. The thick lines
  give fully converged MA results, the dots corresponds to the 3-boson
  solution. The insets show the 3-boson results over a larger
  $\Omega/t$ range, and the dashed lines show $1/x^5$ dependence. See
  text for more details.
  \label{fig9}}
\end{figure}

In Fig.~\ref{fig9} we study the dependence of $t_2, t_3$ and $E_P$ on
$\Omega/t_b$ when $t_f=0$. The full lines show converged MA results
(except for very small $\Omega$, see below), whereas the symbols show
the MA results with the restriction that we only allow clouds with up
to 3 bosons ($A_4=0$). As expected, at large $\Omega$ the agreement is
very good: we know that we need clouds with at least 3 bosons to
generate the effective hoppings, and because $\Omega$ is large, it is
very unlikely to have larger clouds. This is further confirmed by the
insets, which show that in the limit $t_b/\Omega \rightarrow 0$, both
effective hoppings scale like $t_b^6/\Omega^5$, as expected for the
3-boson, 3-site processes from perturbation theory. The fits also
confirm that in this limit, $t_2/t_3=4$. This is because there is
constructive interference in going about the Trugman loops clockwise
and anticlockwise to generate $t_2$, while $t_3$ can only be generated
in a unique way.

\begin{figure}[t]
\includegraphics[width=\columnwidth]{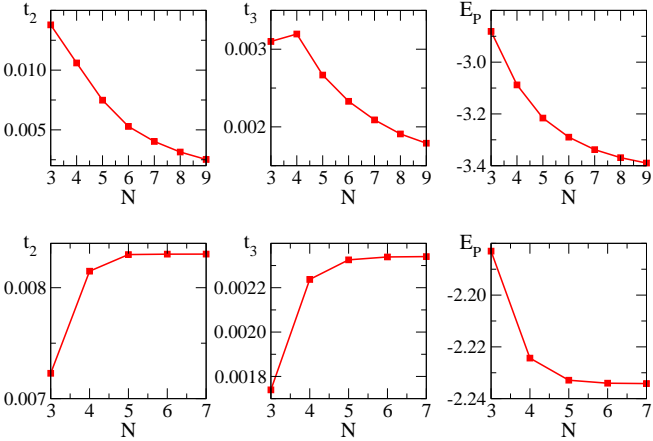}
\caption{ Effective hopping integrals $t_2, t_3$ and
  polaron ground-state energy $E_P$  vs. the
  maximum number of allowed bosons $N$. The upper (lower) panels
  correspond to $\Omega/t_b=0.4$ ($\Omega/t_b=1$). In both cases,
  $t_f=0$.
\label{fig10}}
\end{figure}

As $\Omega$ decreases below roughly $2t_b$, we see that $t_2,t_3$
become, at first, larger than the corresponding 3-boson
values. Indeed, here we have to use larger clouds to achieve full
convergence, and processes with more than 3 bosons will further
increase the effective hoppings. Surprisingly, for $\Omega/t_b<0.7$ or
so, $t_2$ and $t_3$ start to decrease fast. Here many-boson processes
lead to a decrease of the effective hoppings from what the simple
3-boson scenario would predict. Convergence of the various quantities
in dependence on the maximum number $N$ of bosons allowed in the cloud
is shown in Fig.~\ref{fig10} for $\Omega/t_b=0.4$ (upper panels) and
for $\Omega/t_b=1$ (lower panels). The data indeed confirm that 4 and
more boson terms have different effects on the effective hoppings for
$\Omega < t_b$ and $\Omega > t_b$.

It is also clear that in the limit $\Omega/t_b \rightarrow 0$ our
results are untrustworthy, because we ignore longer loops that also
contribute to the effective hoppings in this regime. For example, just
as the 6-step Trugman loop on a $2\times2$ plaquette contributes
to 2nd NN hopping, the Trugman loops on $2\times3$ plaquettes will
contribute to both 2nd and 3rd NN hopping (depending on how the
fermion goes around the loop) and these contributions will supplement
the values obtained from the processes included here.  Its
contributions scale asymptotically like $t^{11}_b/\Omega^{10}$ because
they involve a 5-boson cloud and 11 hoppings to first create and then
annihilate all of them. This is to be contrasted to $t_b^6/\Omega^5$
scaling for the contributions included in the 3-site cloud MA.
Clearly, once $\Omega \sim t_b$ we cannot ignore the contribution of
these longer loops. This is why it is also pointless to go to larger
$N$ to find the fully converged values for the $\Omega/t_b=0.4$ in
Fig.~\ref{fig10}.  However, a good estimate of the crossover is
difficult to obtain from such perturbation theory arguments, since the
insets in Fig.~\ref{fig9} reveal that the asymptotic expressions are
only valid at much larger $\Omega/t_b$ values.

\begin{figure}[t]
\includegraphics[width=\columnwidth]{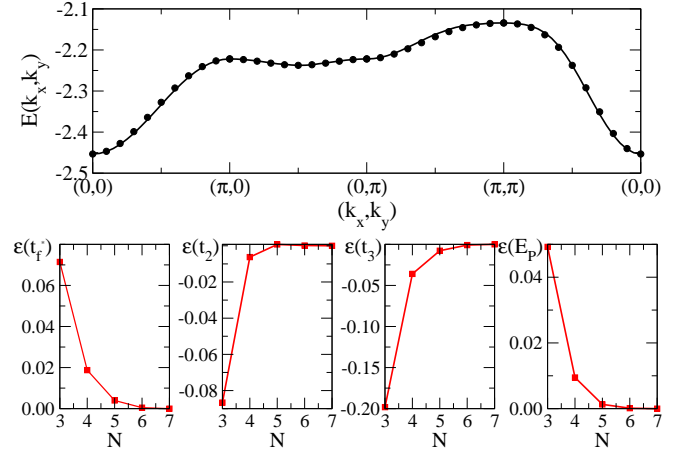}
\caption{Polaron dispersion $E(\kv)$ (upper panel)
along high-symmetry cuts in the 2D square-lattice Brillouin zone, for
$t_f=0.1, t_b=\Omega=1$. Lines show MA results, the symbols are fits
to Eq.~(\ref{27}). The lower panels show the relative error
$\varepsilon(x) = 1- x(N)/x(\infty)$ in the effective hopping
integrals, as well as the polaron ground-state energy, vs. the maximum
number $N$ of allowed bosons.
\label{fig11}}
\end{figure}

A better criterion is to take the value $\Omega/t_b$ above which
convergence is achieved for a cutoff equal to or less than 6,
signaling that 5 or more boson processes are not contributing much to
the polaron wavefunction, and therefore longer loops can be ignored
safely. This definition is not going to produce a very sharp value
since these contributions change gradually. From Fig.~\ref{fig10} we
see that for $\Omega/t_b=1$, the change in going from $N=4$ to $N=5$
modifies various quantities by up to about 4\%, therefore this is
likely already in the regime where longer loops are not playing an
important role.  Once 5-boson processes become important, we need to
include in the variational calculation at least the $3\times 2$ loops
which will modify both $t_2$ and $t_3$. This is why the numbers shown
in Fig.~\ref{fig9} are likely not valid for small $\Omega$. This is an
example of how a MA approximation can signal its potential problems,
but also how to fix them (here, extension to at least 5-site polaron
clouds is needed at lower $\Omega$).

In any case, the boson-modulated hopping is responsible for
dynamically generating a finite effective (dimensionless) mass
$m^* = t_b/(t_2+2t_3)\ge 14$ or so, even though the free particle
has an infinite mass.

We next discuss the case of finite $t_f$, in the limit $t_f< \Omega$
and $\Omega/t_b \ge 1$ where the results of this approximation are
expected to be valid. The main change in the polaron dispersion is
that it will also acquire NN contributions, so now
\begin{eqnarray}
\nonumber E(\kv) &=& E_P -2t_f^*[\cos(k_xa)+\cos(k_ya)-2]\\
\nonumber&&-2t_2[\cos((k_x+k_y)a) + \cos((k_x-k_y)a)-2]\\ \EqLabel{27}
&&- 2t_3[\cos(2k_xa) + \cos(2k_ya)-2]\,,
\end{eqnarray}
where $t_f^*$ is renormalized due to polaron cloud overlap.

\begin{figure}[t]
\includegraphics[width=\columnwidth]{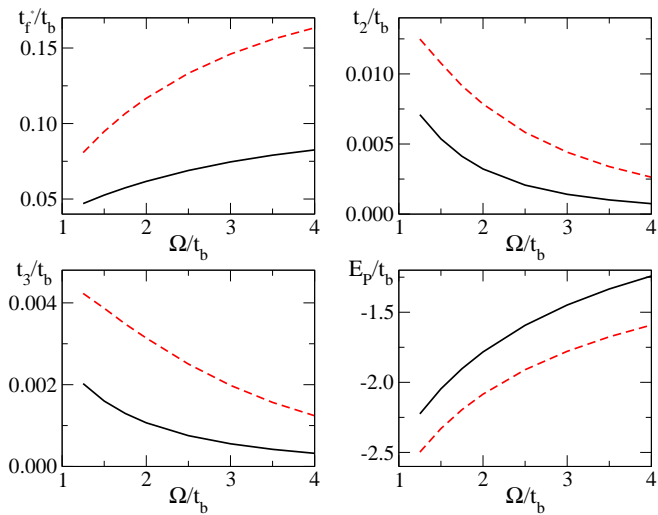}
\caption{(Color online) The effective hopping integrals and the
  polaron energy, in units of $t_b$, vs. $\Omega/t_b$, for
  $t_f=0.1t_b$ (black solid line) and $t_f=0.2 t_b$ (red dashed
  line). The results are only shown for $\Omega$ values where 5 or
  more boson processes become irrelevant.
\label{fig12}}
\end{figure}

The polaron dispersion is shown in Fig.~\ref{fig11} for $t_f=0.1 t_b$,
$\Omega=t_b$. A comparison with Fig.~\ref{fig8} reveals that the small
$t_f$ has a significant effect on $E(\kv)$, especially near the edges
of the Brillouin zone. The symbols show fits to Eq.~(\ref{27}). Here,
the values for $E_P,t_f^*, t_2$ and $t_3$ were extracted from the
energies at the special points $(0,0)$, $({\pi\over2},{\pi\over2})$,
$(0,\pi)$, and $(\pi,\pi)$. Using these parameters in Eq.~(\ref{27})
leads to good agreement with the MA results (thick lines). We find
that $t_f^*/t_b= 0.040, t_2/t_b=0.0090,
t_3/t_b=0.0026$. Interestingly, while $t_f^*<t_f$, as expected in
polaronic physics, we see that in the presence of a finite $t_f$, both
$t_2$ and $t_3$ are larger than when $t_f=0$.  This is because the
effective hopping integrals generated by the 3-boson processes
discussed so far are here supplemented by the usual longer-range
polaron hopping known to occur in the intermediate-to-strong
electron-phonon coupling limit.~\cite{St96,WF97} Just as for $t_f=0$,
the {\em qp} weight changes little throughout the Brillouin zone; it
is close to $0.3$ everywhere. The lower panels in Fig.~\ref{fig11}
show the convergence of the various effective hopping amplitudes and
of the ground-state energy as the maximum number of bosons $N$
increases. Specifically, we plotted the relative errors which reveal
that 5 or more boson processes contribute less than 4\% to the various
quantities.

The dependence of $t_f^*, t_2$, $t_3$ and $E_P$ on $\Omega/t$ is shown
in Fig.~\ref{fig12} for a two values of $t_f$. The data is only
displayed over the range where 5-boson processes contribute less than
1\% to the various parameters, so that longer Trugman loops can be
safely ignored. As expected, $t_f^*$ increases towards $t_f$ as
$\Omega$ increases, because this leads to fewer bosons in the polaronic
cloud, i.e. less ``dressing'' of the quasiparticle. On the other hand,
$t_2$ and 
$t_3$ decrease 
with increasing $\Omega$, as this makes the intermediary many-boson
states less likely. As already discussed, their values increase with
increasing $t_f$, for a fixed value of $\Omega$. Finally, we note that
$E_P$ is well below the free particle continuum starting at
$-4t_f$. This, together with the fact that longer loops are
irrelevant, guarantees that the approximation must be quantitatively
accurate in this regime.

\section{Summary and discussions}

The main goal of this work is to demonstrate how the MA approximations
can be generalized to study polaron formation in models with
boson-affected fermion hopping. Unlike for simpler local fermion-boson
couplings as in the Holstein polaron model, where the strong-coupling
Lang-Firsov solution is known and can guide the choice for the
maximum extension of the polaron cloud, here this solution is
not available. As illustrated for the Edwards fermion-boson model, one
now needs to use physical intuition to make a reasonable choice. Of
course, one can always systematically increase the variational space
and check that the initial guess was indeed reasonable.

Because we are interested primarily in the low-energy polaron physics,
we used the MA$^{(0)}$ level of approximation which describes only the
polaron cloud and does not allow for far-flung bosonic
excitations. Then, the only free ``parameter'' is the spatial size of
the polaronic cloud. Simple arguments regarding the processes
illustrated in Fig.~\ref{fig0} show that at least 3-site clouds need
to be allowed, and therefore we built the approximation for this
case. Careful consideration of the terms ignored gives us intuition
about when the approximation is expected to be accurate, and when 
and in which way it becomes problematic. In 1D, this was indeed
verified successfully against available exact numerical results.

We then extended the calculation to 2D, where no results were
available for this model until now, and demonstrated that the closed
Trugman loops play the key role in determining the effective mass of
the quasiparticle in the limit $t_f=0$. In this regime, the results
for the 3-site MA calculation are trustworthy for $\Omega > t_b$; for
smaller $\Omega$ values, one needs to increase the allowed size of the
polaron cloud since longer Trugman loops are also becoming
important. We emphasize that the MA approximation is not wrong in the
low $\Omega$ limit; what failed is our guess about the relevant size
of the polaron cloud. If this is increased from 3 to more sites, the
approximation will become accurate again.

The important role played by Trugman loops raises a very important
question regarding the motion of a particle through an AFM background
(which, as discussed,  the Edwards fermion-boson model partly
mimics). For $t-J$ models, it has been
argued that the interaction of the hole with spin-waves is well
described within the self-consistent Born approximation. This
approximation includes only non-crossing diagrams, i.e. processes
where the bosons are absorbed in inverse order to the one in which
they have been emitted by the particle. This is because it is
generally expected that the particle needs to retrace its path to
``heal'' the string of defects it created when it reshuffled the
spins. For a Ne\'el AFM, the self-consistent Born approximation
therefore predicts an infinitely heavy quasiparticle. In the presence
of spin-fluctuations, the magnons can disperse and this gives rise to
a finite quasiparticle mass.

What is shown here is that the quasiparticle can acquire a finite spin
mass even in the absence of spin fluctuations, by going (almost) twice
around closed loops, first creating a string of defects and then
healing it. Note that in diagrammatic terms, this would correspond to
maximally crossed diagrams, since here the bosons are absorbed in the
same order in which they were emitted. Such processes are obviously
not included in the  self-consistent Born approximation.

Of course, one might argue that spin fluctuations are relevant for
holes in cuprates, in other words $t_f$ may be considerable and
therefore dominate $E(k)$, as we found it to be the case for the
Edwards model for $t_f>0$. In other words, that the Trugman loops'
contributions, although finite, may be quantitatively
insignificant. However, the interesting thing is that these
closed loop processes give rise to the only contributions in $E(k)$
which are consistent with the doubling of the unit cell.  ARPES on the
parent insulators, i.e. for a single quasiparticle introduced in  a Cu0$_2$
layer,   clearly exhibits this
doubling of the Brillouin zone.\cite{ARPES}  This suggests that maybe this
problem should 
be revisited using MA-type approximations.

\acknowledgements The authors would like to thank A. Alvermann and
D. M. Edwards for valuable discussions. This work was supported by
NSERC and CIfAR (MB), and by Deutsche Forschungsgemeinschaft
through SFB 652 (HF).



\end{document}